\documentclass
[twocolumn,aps,prd,amsmath,showpacs,amssymb,nofootinbib]
{revtex4}

\usepackage[dvips]{graphicx}
\usepackage{bm}                      
\usepackage{mathptmx}                
\usepackage{dcolumn}                 

\def\bea{\begin{eqnarray}}
\def\eea{\end{eqnarray}}
\def\be{\begin{equation}}
\def\ee{\end{equation}}

\def\P3{{\cal P}_t}
\def\J3{{\cal J}}
\def\T3{{\cal T}}

\def\bra{\langle}
\def\ket{\rangle}

\def\egu{\, =\, }

\def\cap{\noindent}

\def\beq{\begin{equation}}
\def\eeq{\end{equation}}
\def\bar{\begin{array}[b]}
\def\barc{\begin{array}}
\def\bart{\begin{array}[t]}
\def\ear{\end{array}}
\def\le#1{\label{eq:#1}}

\begin{document}

\title{Selecting microscopic Equations of State}

\author{G. Taranto$^1$, M. Baldo$^1$, and G. F. Burgio$^1$}

\affiliation{
$^1$INFN Sezione di Catania, and Dipartimento di Fisica e Astronomia,
Universit\'a di Catania, Via Santa Sofia 64, 95123 Catania, Italy}


\begin{abstract}
We compare a set of equations of state derived within microscopic many-body approaches,
and study their predictions as far as phenomenological data on nuclei from heavy ion collisions,
and astrophysical observations on neutron stars are concerned. We find that all the data, taken together, put 
strong constraints not easy to be fulfilled accurately.  However, no major discrepancies are found
among the selected equations of state and with respect to the data. The results provide an
estimate of the uncertainty on the  theoretical prediction at a microscopic level of the nuclear equation of state. 
\end{abstract}


\maketitle

\section{Introduction}

 A convergent effort of experimental and theoretical nuclear physics has been developing along several years to determine or to constrain the Equation of State (EoS) of nuclear matter.
Two main areas of research have provided relevant hints in this direction. The experimental data on heavy ion collisions have been systematically analyzed on the basis of detailed 
simulations, in order to constrain not only the density dependence of the EoS, but also its isospin dependence. Excellent reviews on this subject can be found in the literature \cite{aich,li}.
 On the other hand, the analysis of astrophysical observational data, noticeably on compact objects,
are of great relevance for the study of the nuclear EoS.
The results in this field of research can be considered complementary to the ones that can be
obtained within the heavy ion research activity, because nuclear matter is involved
in different physical conditions. In Neutron Stars (NS) nuclear matter is present in beta equilibrium from
very low density to several times saturation density, and it is therefore extremely
asymmetric, much more than nuclei in laboratory. Despite the different physical situations, 
an accurate microscopic theory of nuclear matter  is expected to be able to explain correctly 
the data obtained in both physical realms.  For a pedagogical introduction to the field see ref. \cite{bb}.
Another more phenomenological approach is based on the Energy Density Fuctional (EDF) method, in particular the Skyrme forces scheme. In this approach a phenomenological force or EDF
includes a certain set of parameters that are fixed by fitting the binding energy of nuclei
throughout the mass table. Some of these forces are adjusted also to microscopic nuclear matter
EoS. Recently, an ample set of Skyrme forces, that have been presented in the literature, has been
analyzed and confronted with the available constraints on the nuclear EoS obtained 
from heavy ion reactions and astrophysical objects \cite{dutra}. The few Skyrme forces that passed 
these tests have been then used to predict the binding energy of a wide set of nuclei \cite{stev},
and none seems to perform satisfactorily well in this case.
These results show clearly that it is not at all trivial to satisfy the constraints coming
from astrophysical and heavy ion data, as it was already found in ref.\cite{klahn}. Furthermore, if the nuclear mass data are included, it seems
very difficult to reproduce all the data sets. It looks that the phenomenological forces are
not flexible enough to this purpose.   In this paper we follow a complementary  approach. We focus on the microscopic theory of nuclear matter and the corresponding  EoS's that have been developed so far.  Adopting the same line as in ref.\cite{dutra}, we test the predictions of different
microscopic many-body approaches with respect to the constraints coming from
experiments and phenomenology. The aim of this analysis is to establish to what 
extent the data can be reproduced within a microscopic approach and the accuracy  
of the present theory of nuclear matter.     

This paper is organized as follows. In Sect.II a review of the currently used microscopic
many-body methods is presented, whereas results are discussed in Sect. III.  Finally, in Sect. IV we draw our conclusions.

\bigskip

\section{EoS of nuclear matter  in microscopic approaches}
\label{s:EOS}

Empirical properties of infinite nuclear matter can be calculated using many different theoretical approaches. In this paper we concentrate on the microscopic ones, the only input being a realistic free nucleon-nucleon (NN) interaction with parameters fitted to NN scattering phase shifts in different partial wave channels, and to properties of the deuteron. In the following we discuss 
the non-relativistic Brueckner-Hartree-Fock (BHF) method \cite{book} and its relativistic counterpart,
the Dirac-Bruckner- Hartree-Fock (DBHF) approximation \cite{Machleidt89}, and the Variational 
method \cite{PandaWir}. 
A relatively more recent approach is based on the Chiral effective Field Theory, which has
been extensively applied to Nuclear Matter. For a review see ref. \cite{Oller,Meissr}. It establishes
a link with the underlying QCD structure of strong interaction, and it has reached
a high degree of sophistication \cite{Meiss,Hebel,Weise1,Weise2,Mach}. We will not
discuss this approach, because the corresponding EoS is based on a low momentum expansion,
and therefore its behaviour at high density is uncertain. 
Hence, a comparison with the other microscopic EoS's would be incomplete. However, 
the EoS from the Chiral approach turns out to be close
to the variational EoS \cite{Oller}, and therefore the parameters near saturation are expected to
be quite similar.

\subsection{The Brueckner-Bethe-Goldstone approach}

The Brueckner--Bethe--Goldstone (BBG) theory is based on a linked cluster 
expansion of the energy per nucleon of nuclear matter (see Ref.~\cite{book},
chapter 1 and references therein).  
The basic ingredient in this many--body approach is the Brueckner reaction 
matrix $G$, which is the solution of the  Bethe--Goldstone equation 

\begin{equation}
G[\rho;\omega] = v  + \sum_{k_a k_b} v {{|k_a k_b\rangle  Q  \langle k_a k_b|}
  \over {\omega - e(k_a) - e(k_b) }} G[\rho;\omega], 
  \label{eq:G}
\end{equation}                                                           
\noindent
where $v$ is the bare NN interaction, $\rho$ is the nucleon 
number density, and $\omega$ the  starting energy.  
The single-particle energy $e(k)$ (assuming $\hbar$=1 here and throughout 
the paper),
\begin{equation}
e(k) = e(k;\rho) = {{k^2}\over {2m}} + U(k;\rho),
\label{e:en}
\end{equation}
\noindent
and the Pauli operator $Q$ determine the propagation of intermediate 
baryon pairs. The Brueckner--Hartree--Fock (BHF) approximation for the 
single-particle potential
$U(k;\rho)$  using the  {\it continuous choice} is
\begin{equation}
U(k;\rho) = {\rm Re} \sum _{k'\leq k_F} \langle k k'|G[\rho; e(k)+e(k')]|k k'\rangle_a,
\end{equation}
\noindent
where the subscript ``{\it a}'' indicates antisymmetrization of the 
matrix element.  
Due to the occurrence of $U(k)$ in Eq.~(\ref{e:en}), they constitute 
a coupled system that has to be solved in a self-consistent manner
for several  momenta of the particles involved, at the considered densities.
In the BHF approximation the energy per nucleon is
\begin{equation}
{E \over{A}}  =  
          {{3}\over{5}}{{k_F^2}\over {2m}}  + {{1}\over{2\rho}}  
~ \sum_{k,k'\leq k_F} \langle k k'|G[\rho; e(k)+e(k')]|k k'\rangle_a. 
\end{equation}
\noindent
In this scheme, the only input quantity we need is the bare NN interaction
$v$ in the Bethe-Goldstone equation (1).  The nuclear EoS can be 
calculated with good accuracy in the Brueckner two hole-line 
approximation with the continuous choice for the single-particle
potential, since the results in this scheme are quite close to the 
calculations which include also the three hole-line
contribution \cite{song}. 

However, it is commonly known that  
non-relativistic calculations, based on purely two-body interactions, fail 
to reproduce the correct saturation point of symmetric nuclear matter.
One of the well known results of several studies, that lasted for 
about half a century, is the need of introducing three-body forces (TBFs).
In our approach the TBF is reduced to a density dependent two-body force by
averaging over the position of the third particle, assuming that the
probability of having two particles at a given distance is reduced 
according to the two-body correlation function \cite{bbb}.

In this work we will illustrate results for two different approaches to the TBF's, 
i.e. a phenomenological  and a microscopic one. The phenomenological approach is based on  the so-called Urbana model,  which consists of an attractive term due to two-pion exchange
with excitation of an intermediate $\Delta$ resonance, and a repulsive 
phenomenological central term \cite{uix}. We introduced the same Urbana three-nucleon model within the BHF approach.  Those TBF's produce a shift of about $+1$ MeV in energy and $-0.01$ fm $^{-3}$ in density. This adjustment is obtained by tuning the two parameters
contained in the TBF's,  and was performed to get an optimal saturation point (the minimum) (for details see Ref.~\cite{bbb}).

The connection between two-body and three-body forces within the meson-nucleon theory of
nuclear interaction is extensively discussed and developed in references \cite{tbfmic,zhli}. 
At present the theoretical status of microscopically derived TBF's is still quite rudimentary, however a tentative approach has been proposed using the same meson-exchange parameters as the 
underlying NN potential. Results have been obtained with the Argonne $v_{18}$, the Bonn B, and the Nijmegen 93 potentials \cite{zhli}. 
 
In the past years, the BHF approach has been extended in order to include the hyperon degrees of freedom
\cite{sch98}, which play an important role in the study of neutron star matter. However, in this paper we are mainly interested on the properties of the nucleonic EoS, therefore this issue will not be discussed further.

\subsection{The Relativistic approach}
The relativistic framework is the one on which the nuclear EoS should be
ultimately  based. The best relativistic treatment developed so far is the Dirac-Brueckner (DBHF) approach \cite{Machleidt89}. The DBHF method can be developed in analogy
with the non-relativistic case, i.e. the nucleon inside the nuclear medium is viewed
as a dressed particle in consequence of its two-body interaction with the surrounding
nucleons. The two-body correlations are described by introducing the in-medium
relativistic $G$-matrix. The DBHF scheme can be formulated as a self-consistent problem between the single
particle self-energy $\Sigma$ and the $G$-matrix. Schematically, the equations can be written
\begin{eqnarray}
G  &=&  V + i\int V Q g g G  \\
\Sigma &=& -i \int_F(Tr[gG] - gG) \label{eq:sig}
\end{eqnarray}
\noindent where $Q$ is the Pauli operator which projects the intermediate two particle momenta outside the Fermi
sphere, as in the BBG G-matrix equation (\ref{eq:G}), and $g$ is the single particle Green's function,
which fulfills the Dyson equation
\begin{equation}
g = g_0 + g_0 \Sigma g
\end{equation}
\noindent 
where $g_0$ is the (relativistic) single particle Green's function for a free gas of nucleons,
and $\Sigma$ is the nucleon self-energy which expresses the influence of the surrounding nucleons. The self-energy can be expanded in the covariant form
\begin{equation}
\Sigma(k,k_F) = \Sigma_s(k,k_F) - \gamma_0\Sigma_0(k,k_F) + \mbox{\boldmath $\gamma$} \cdot{\bf k}\Sigma_v
\label{eq:sigex}
\end{equation}
\noindent  
where $\gamma_\mu$ are the Dirac gamma matrices, and the coefficients of the expansion are scalar
functions, which in general depend on the modulus $ |{\bf k}| $ of the three-momentum and on the energy $k_0$. The free single particle eigenstates, which determine the spectral representation of the free Green's function, are solutions of the Dirac equation
\begin{equation}
[\,\,\, \gamma_\mu k^\mu \, - \, M\,\,\, ]\, u(k)\,\, =\, 0
\end{equation}
\noindent where $u$ is the Dirac spinor at four-momentum $k$. For the full single particle Green's function $g$
the corresponding eigenstates satisfy
\begin{equation}
[\,\,\, \gamma_\mu k^\mu \, - \, M \, + \, \Sigma \,\,\,]\, u(k)^*\,\, =\, 0
\end{equation}
\noindent 
Inserting the above general expression for $\Sigma$, eq.(\ref{eq:sigex}), after a little manipulation one gets
\begin{equation}
[\,\,\, \gamma_\mu {k^\mu}^* \, - \, M^*\,\,\, ] u(k)^*\,\, =\, 0
\end{equation}
\noindent  with 
\beq
 {k^0}^* \,=\, {k^0 + \Sigma_0\over 1 + \Sigma_v} \,\,\,\,\,\, ;\,\,\,\,\,\, {k^i}^* \,=\,
 k^i
 \,\,\,\,\,\, ; \,\,\,\,\,\, M^* \,=\, {M + \Sigma_s\over 1 + \Sigma_v}
 \label{eq:momen}
\eeq \noindent 
This is the Dirac equation for a single particle in the medium, and the corresponding solution is
the spinor
\begin{equation} 
{u}^*({\bf k},s)=\sqrt{\frac{{E}^*_{\bf k}+{M}^*}{2 {M}^*}}
\left( \begin{array}{c} 1\\ \frac{\mbox{\boldmath $\sigma \cdot k$}}{{E}_{\bf k}^*+{M}^*}
\end{array} \right) \chi_{s}  \,\,\, ;  \,\,\, {E}^*_{\bf k} = \sqrt{{\bf
k}^2 + {M^*}^2 } \,\,\,  . \label{eq:spino}
\end{equation}
In line with the Brueckner scheme, within the BBG expansion, in the
self-energy of equation (\ref{eq:sig}) only the contribution of the single particle Green's function pole is
considered. Furthermore, negative energy states are neglected and one gets the usual
self--consistent condition between self--energy and scattering $G$--matrix. 
 
In any case, the medium effect on the spinor of equation
(\ref{eq:spino}) is to replace the vacuum value of the nucleon mass and three--momentum with the in--medium values
of equation (\ref{eq:momen}). This means that the in--medium Dirac spinor is ``rotated" with respect to the
corresponding one in vacuum, and a positive (particle) energy state in the medium has some non--zero component on
the negative (anti--particle) energy state in vacuum. In terms of vacuum single nucleon states, the nuclear medium
produces automatically anti--nucleon states which contribute to the self--energy and to the total energy of the
system. It has been shown in ref. \cite{Brown87} that this relativistic effect is equivalent to the
introduction of well defined TBF's at the non--relativistic level. These TBF's turn out to be repulsive, and consequently produce a saturating effect.  Actually, including in BHF only these particular TBF's, one gets results close to DBHF calculations, see ref.\cite{Compilation}.
Generally speaking, the DBHF gives in general a better saturation point  than BHF,  
and the corresponding EoS turns out to be stiffer above saturation than the
one calculated from the BHF + TBF method.

In the relativistic context the only NN potentials which have been developed are the ones of one boson exchange (OBE) type. In the calculations shown here the Bonn A potential is used \cite{fuchs99}.

\subsection{The Variational method}

In the variational method \cite{PandaWir} one assumes that the ground state wave function $\Psi$ can be written in
the form 
\be
     \Psi(r_1,r_2,......) \, =\, \Pi_{i<j} f(r_{ij}) \Phi(r_1,r_2,.....)
     \, ,
     \label{eq:corr}
 \ee 
\noindent 
where $\Phi$ is the unperturbed ground state wave function, properly antisymmetrized,
and the product runs over all possible distinct pairs of particles. The correlation factor is here determined by the variational principle, i.e. by imposing that the mean value of the Hamiltonian gets a minimum 
\be
   {\delta\over \delta f} { {\bra \Psi \vert H \vert \Psi \ket }\over
   {\bra \Psi \vert \Psi \ket} } \,= \, 0 \,. 
\ee  \noindent 
In principle this is a functional equation for the correlation function $f$, which
however can be written explicitly in a closed form only if additional suitable approximations are introduced. The
function $f(r_{ij})$ is assumed  to converge to $1$ at large distance and to go rapidly to zero as $r_{ij}
\rightarrow  0$, to take into account  the repulsive hard core of the NN interaction. Furthermore, at distance
just above the core radius a possible increase of the correlation function beyond the value $1$ is possible.\par
For nuclear matter it is necessary to introduce a channel dependent correlation factor, which is equivalent to
assume that $f$ is actually a two-body operator $\hat{F}_{ij}$. One then assumes that $\hat{F}$ can be expanded in
the same spin-isospin, spin-orbit and tensor operators appearing in the NN interaction. Momentum dependent
operators, like spin-orbit, are usually treated separately. The product in  equation (\ref{eq:corr}) must be then
symmetrized since the different terms do not commute anymore.
\par
If the two-body NN interaction is local and central, its mean value is directly related to the pair distribution
function $g({\bf r})$ \be
 < V > \, =\,  {1\over 2}\rho \int d^3r v(r) g({\bf r})  \, ,
\ee \noindent 
where 
\be
 g({\bf r_1 - r_2}) \, =\, {\int \Pi_{i>2}d^3r_i \vert\Psi(r_1,r_2....)\vert^2
    \over  \int \Pi_{i}d^3r_i \vert\Psi(r_1,r_2....)\vert^2  } \, .
\ee
\par
The main job in the variational method is to relate the pair distribution function to the correlation 
factors $F$.
Again, in nuclear matter also the pair distribution function must be considered channel dependent and the relation
with the correlation factor becomes more complex. In general this relation cannot be worked out exactly, and one
has to rely on some suitable expansion. Furthermore, three-body or higher correlation functions must in general be
introduced, which will depend on three or more particle coordinates and describe higher order correlations in the
medium. Many excellent review papers exist in the literature on the variational method and its extensive use for
the determination of nuclear matter EoS \cite{Navarro,PandaWir}. The best known and most used variational nuclear
matter EoS is the one calculated by Akmal, Pandharipande and Ravenhall \cite{akma}. In their paper
the authors showed calculations using the Argonne $v_{18}$ NN interaction \cite{v18}, with
boost corrections to the two-nucleon interaction, which give the leading relativistic effect of 
order $(v/c)^2$, as well as three-nucleon interactions modeled with the Urbana force explained above. This EoS will be tested in the present paper.

\begin{table*}[t]
\begin{center}
\begin{tabular}{|l|c|c|c|c|c|}
\hline
 EoS &  $\rho_0 \rm (fm^{-3})$ & $\frac{E}{A}$ (MeV)& $K_0$ (MeV)& $S_0 $ (MeV) & $L$ (MeV)\\
\hline\hline
$BHF, \, Av_{18} + UVIX$  & 0.16 & -15.98 & 212.4 & 31.9  & 52.9  \\
\hline
$ BHF, \, Av_{18} + micro~ TBF $ & 0.2 & -15.5 & 236. & 31.3 & 82.7  \\
\hline
$ BHF, \, Bonn~ B + micro~ TBF $ & 0.17 & -16. & 254. & 30.3 & 59.2  \\
\hline
$ APR,  \, Av_{18} + UVIX$ & 0.16 & -16. & 247.3 & 33.9 & 53.8  \\
\hline
$ DBHF,  \, Bonn~ A $ & 0.18 & -16.15 & 230. & 34.4 & 69.4  \\
\hline
\end{tabular}
\end{center}
\caption{Calculated properties of symmetric nuclear matter.}
\label{t:sat}
\label{table}
\end{table*}  

\section{Results and discussion}

\subsection{Phenomenology of nuclei}
One of the main goals of the microscopic many-body methods is the correct
reproduction of the nuclear matter saturation point.  The non-relativistic  approaches typically 
lead to an over prediction of the saturation density $\rho_0 = 0.17 \pm 0.03 fm^{-3}$  
of symmetric nuclear matter (SNM),  at which the binding energy $E/A$ per nucleon 
reaches its minimum. The empirical value $E/A(\rho_0) \approx - 16~\rm MeV$ can be extracted from the semi-empirical mass formula or from the extrapolation of binding energies of heavy nuclei. 
In Table \ref{t:sat} we display theoretical calculations of saturation properties of SNM for the 
different approaches illustrated above. 

In the case of the BHF scheme with the Urbana three-body forces (first line of Table I), the reported values for the saturation point have been obtained by a polynomial fit to the numerically calculated EoS. The fit was performed with the constraint to reproduce the empirical saturation density of $0.16$ fm$^{-3}$ while the corresponding energy was tuned to give the optimal value within a recent  Energy Density Functional (EDF) approach based on this BHF EoS \cite{BCPM}. The shifts in the values for the density and energy at saturation from the original numerical values (0.17 fm$^{-3}$ and $-15.2$ MeV ) have only marginal effects on the other physical parameters reported in Table I. 
For the BHF case with microscopic three-body forces (second and third lines), no application to an EDF scheme was performed, so the reported saturation point corresponds to the numerical calculations.  
\par
In the case of the considered variational method (fourth line), in the quoted reference the original numerical EoS was corrected around saturation with an extra binding  energy in functional form in order to reproduce the empirical saturation point, as reported in Table I. This correction is large as much as 4 MeV.
We do not know how much this correction could affect the values of the other physical parameters reported in the table, but we used the corrected EoS, since it is the one recommended in the original paper \cite{akma} and widely used in the literature. In particular it was used to construct the   functional NRAPR \cite{Steiner} and in many astrophysical studies on Neutron Stars, see e.g. ref. \cite{hhj}.       
Finally, in the DBHF approach there is no need to introduce three-body forces, 
although the saturation point is slightly shifted to higher density. We therefore report the calculated values without fine tuning or corrections. \par
It has to be stressed that in all cases the EoS above saturation was in no way corrected. 

In the third column of the same table we show the values predicted for the incompressibility
at the saturation point.  For practical reasons, it is customary among nuclear physicists to use the
definition 
\beq 
K_0 \egu k_F^2\left({d^2E/A \over dk_F^2}\right)_{\rho_0} \le{comn} 
\eeq 
\cap 
which has the dimension of an energy. The values of the incompressibility extracted from the
EoS in the many-body methods are displayed in Table \ref{t:sat}. The values are typical of a 
pretty soft EoS at saturation, in agreement with the values extracted from the phenomenology
of monopole oscillations \cite{blai}. One finds indeed that a correlation exists between incompressibility and
position of the monopole Giant Resonance, so that, in principle it is possible to extract from the experimental data the value of the incompressibility in nuclear matter. 
At present, the constraints on the value of the nuclear matter incompressibility from the
monopole excitation are not so tight. It can be approximately constrained between 210 and 250 MeV
\cite{dutra}, though a more refined value can be expected to come out in the near future from additional analysis of phenomenological data.\par 

These values are compatible with the ones obtained from the experiments on sub-threshold kaon production, noticeably the ones from the KaoS \cite{kaos} and FOPI \cite{fopi} collaborations.
The optimal energy for this type of investigation is close or even below two-body threshold, since then the only way to produce the kaons is by compression of the matter. Since at threshold the production rate increases steeply, there is a strong sensitivity to the value of the maximum density reached during the collision, and this is an ideal situation for studying the EoS and its incompressibility. The comparison of the simulations with the experimental data on $K^+$ production  points in the direction of a soft EoS. However it has to be kept in mind that, in the simulations, kaon production occurs at density  $\rho \geq 2-3\rho_0$, and therefore this set cannot be directly compared to the one of the monopole oscillations. In
any case, a stiff EoS above saturation seems to be excluded from this analysis, as it is apparent in ref. \cite{Fuchs}.

Further important properties characterizing the nuclear matter are the symmetry energy at saturation
$S(\rho_0)$ and its density derivative $L$, which are displayed in columns 4 and 5 of 
Table \ref{t:sat}. For both those quantities experimental data do exist, and will be discussed in the 
next subsections.
The usual way of calculating them is to assume asymmetric nuclear matter, where the proton number $N_p$ is different from the neutron number $N_n$, with $\, N = N_n + N_p\, $. In this case,
the EoS of nuclear matter has to be generalized. Defining 
\beq
   \beta \egu {{N_n - N_p}\over {N_n + N_p}} \egu
    {{\rho_n - \rho_p}\over \rho}
\le{as} \eeq \cap as the ``asymmetry'' parameter, one easily gets 

\begin{eqnarray}
\frac{E}{A}(\rho, \beta) &=& \frac{E}{A}(\rho, 0) + E_{sym}(\rho) \beta^2  \\ 
& = &  \frac{E}{A}(\rho_0) + \frac{1}{18} K_0 \, e^2 + \Big[S_0 - \frac{1}{3} Le + \frac {1}{8} K_{sym} e^2 \Big ] \beta^2 \nonumber
\end{eqnarray}
where $e = (\rho - \rho_0)/\rho_0$, $K_0$ is the incompressibility at the saturation point,
$S_0=S(\rho_0)$ is the symmetry energy coefficient at saturation and the parameters $L$ and
$K_{sym}$
in the square bracket characterize the density dependence of the symmetry energy around the 
saturation point. 
The values in Table I of these physical parameters are obtained, for each one of  the indicated microscopic many-body EoS, from their analytical form or by a polynomial fit to the numerical values reported in the original papers.

\begin{figure}[t]
\centering
\includegraphics[scale=0.52,clip]{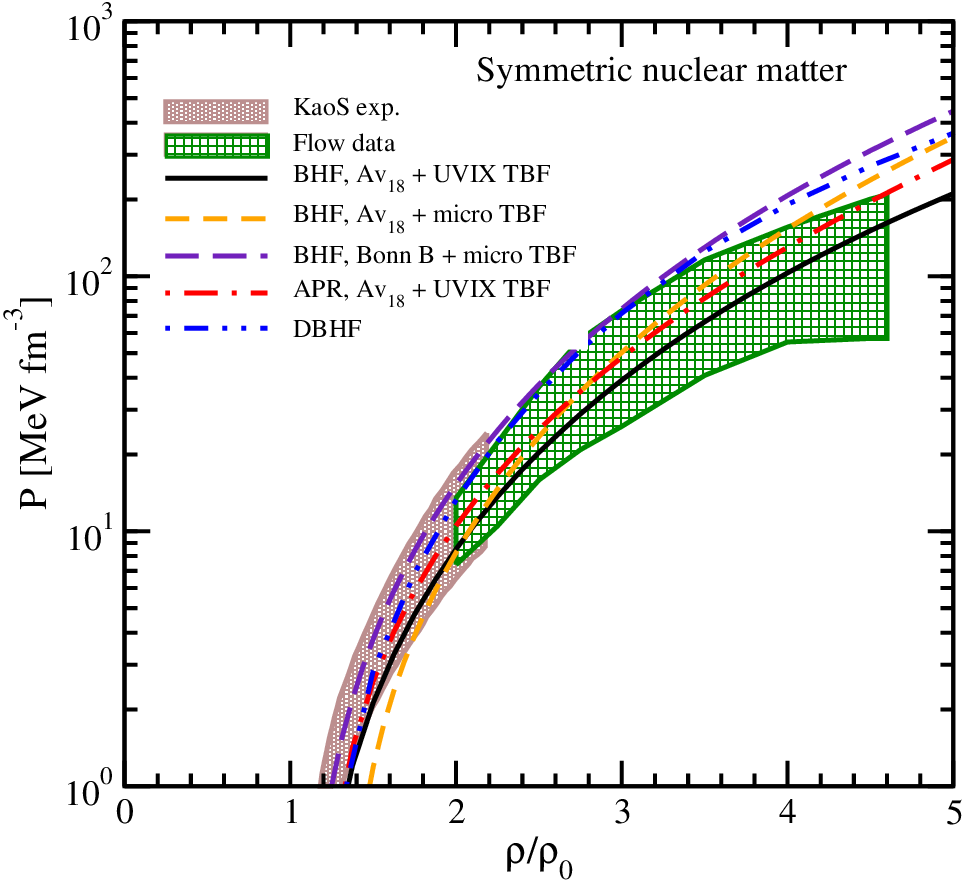}
\caption{(Color online)
Pressure as a function of baryon density for symmetric nuclear matter.
See text for details.}
\label{fig:p_sym}
\end{figure}

\begin{figure}[t]
\includegraphics[scale=0.5,clip]{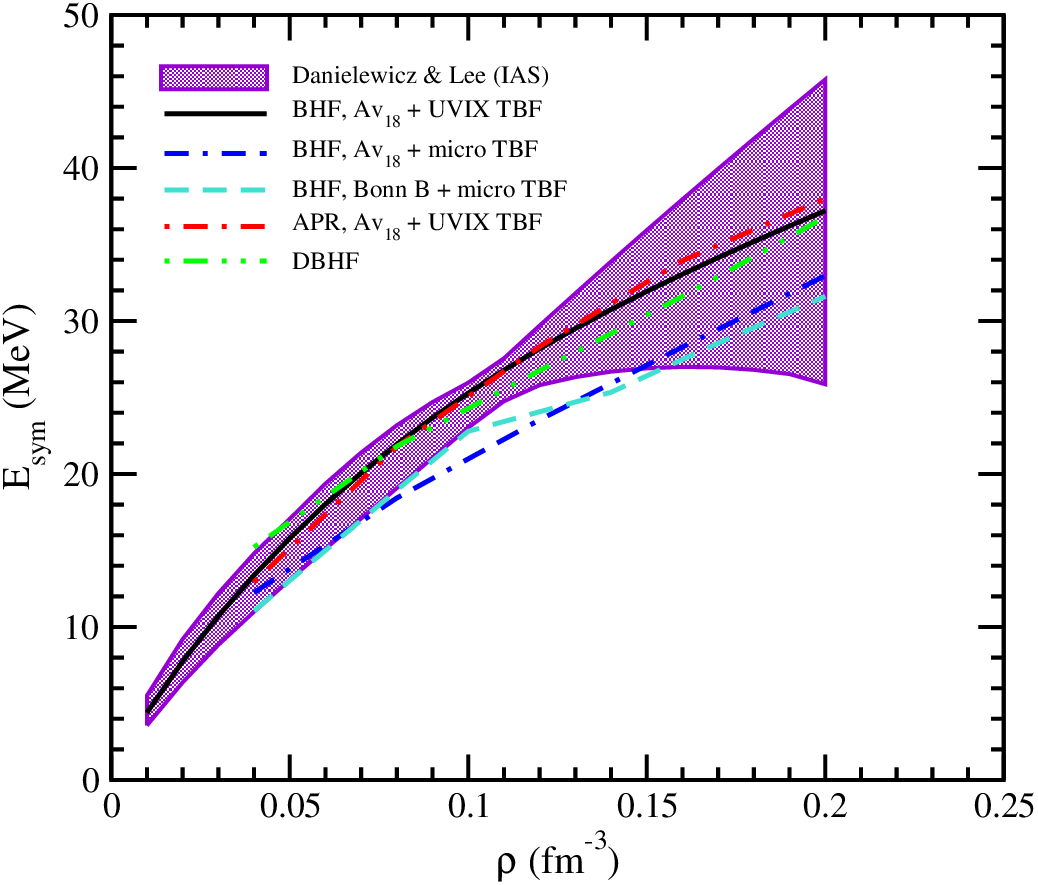}
\caption{(Color online)
The symmetry energy is displayed as a function of the nucleon density.
The purple zone represents the recent data by P. Danielewicz, whereas
the different curves are the results of the microscopic many-body methods.}
\label{f:esym}
\end{figure}

\subsection{Heavy ions phenomenology}

In the last two decades intensive studies of heavy ion reactions at energies
ranging from few tens to several hundreds MeV per nucleon (hereafter indicated as MeV/A) have been performed. The main goal has been the extraction from the data of the gross properties of the nuclear EoS. It can be expected that in heavy ion collisions at large enough energy nuclear matter is compressed and that, at the same time, the two partners of the collisions produce flows of matter. In principle the dynamics of the collisions should be connected with the properties of the nuclear medium EoS and its viscosity. In the so called ''multifragmentation" regime,  after the collision
numerous nucleons and fragments of different sizes are emitted, and the transverse flow, which is strongly affected by the matter compression during the collision, can be measured.

Based on numerical simulations, in reference \cite{DL} a phenomenological range of densities
was proposed where any reasonable EoS for symmetric nuclear matter should pass through in the pressure vs. density plane. 
The plot is reproduced in Fig.\ref{fig:p_sym}, where a comparison with the microscopic 
calculations discussed in Section \ref{s:EOS} is made. The green dashed box represents the
results of the numerical simulations of the experimental data discussed in ref.\cite{DL},
and the brown filled region represents  the experimental data on kaons production \cite{Fuchs}.
We notice that the EoS calculated with the BHF and the variational methods including UVIX three-body forces  look in agreement with the data in the full density range. On the other hand, the
BHF EoS obtained using microscopic TBF's are only marginally compatible with the
experimental data, as well as the DBHF EoS, showing that they are too repulsive already at density
$\rho \geq 3 \rho_0$ if the Bonn potentials are used. Though, it has to be stressed that all EoS are compatible with 
the data around the saturation density, i.e. their incompressibility is as soft as required 
by the data. However, the values of the incompressibility do not characterize completely the EoS, since it is density dependent, but in any case the analysis indicates that the EoS al low density must be soft.

A further constraint on the EoS is given by the symmetry energy, which has been extensively studied 
both from the theoretical and experimental point of view in ref.\cite{tsang2012}. The symmetry 
energy is displayed in Fig.\ref{f:esym}  as a function of the nucleon density. The purple region  is the result of a recent analysis
performed by P. Danielewicz on the isobaric analog states (IAS) in nuclei \cite{iasdan}.
This stems from the charge independence of nuclear interactions, i.e. strong interactions
between nucleons in the same state  do not depend on whether the nucleons are
protons or neutrons. Therefore the energy difference between the ground state of a nucleus
with $N>Z$  and the isobaric analogs of the ground states of neighboring isobars are given by the
symmetry energy, and the Coulomb contributions to the binding energy can be
determined using the IAS. Many such states have been identified , and by fitting the
available data on the IAS, Danielewicz and Lee obtained the constraint shown as a 
purple region in Fig.\ref{f:esym}.
We observe that all EoS give results in very good agreement with the experimental data, except 
the ones of BHF with microscopic TBF at densities below the saturation density.

\begin{figure}[t]
\hspace{-10mm}
\includegraphics[scale=0.5,angle=0, clip]{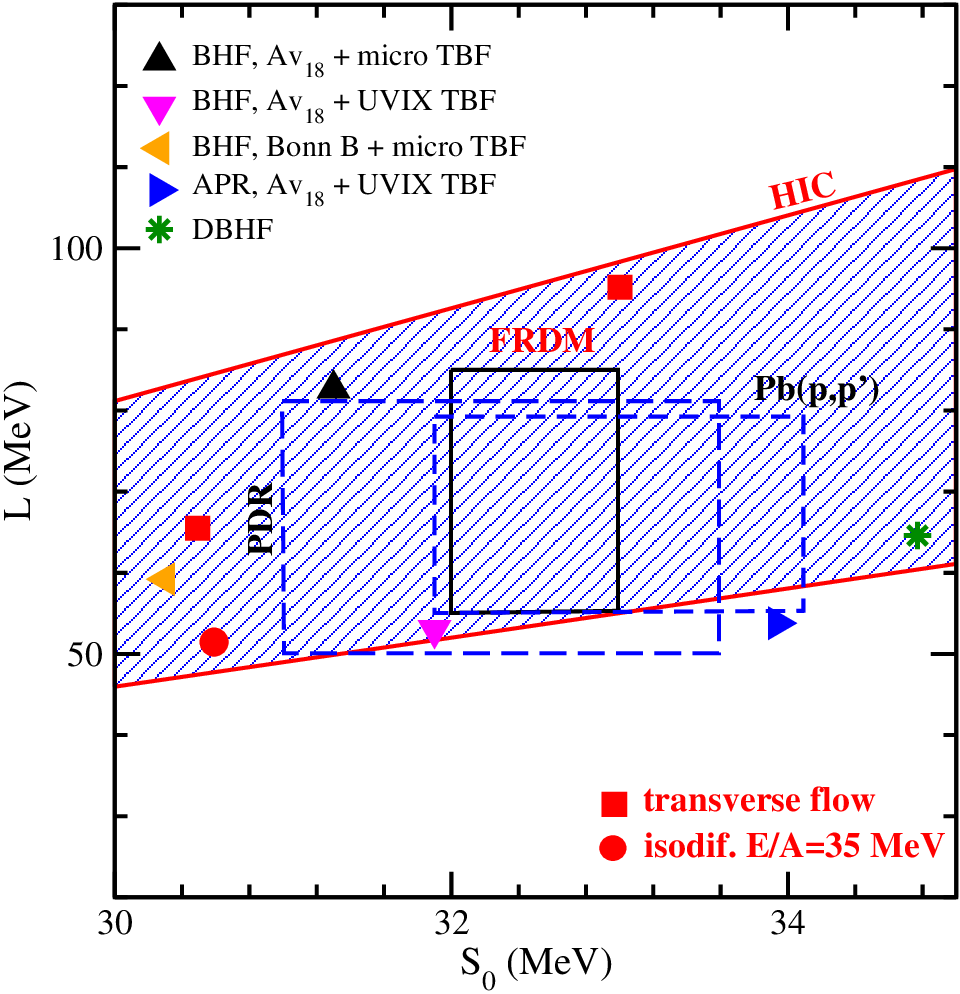}
\caption{(Color online)
The derivative of the symmetry energy $L$ is shown as a function of the symmetry energy
at saturation $S_0$. See text for details.}
\label{f:L}
\end{figure}

In Fig.\ref{f:L} we display the slope L as a function of the symmetry energy at saturation density,
which has been widely discussed in ref.\cite{tsang2012}. Several experimental data are displayed. The blue band represents experimental data 
from HIC, obtained from the neutron and proton spectra from central collisions for
$\rm ^{124}Sn + ^{124}Sn$ and $\rm ^{112}Sn + ^{112}Sn$ reactions at 50 MeV/A \cite{fam}.
At the same incident energy, isospin diffusion was investigated. We remind that isospin
diffusion in HIC depends on the different $N/Z$ asymmetry of the involved projectiles and targets,
hence it is used to probe the symmetry energy \cite{isodif1,isodif2,li}. 
The full red circle shows the results from isospin
diffusion observables measured for collisions at a lower beam energy of 35 MeV per nucleon \cite{isostar}.

Transverse collective flows of hydrogen and helium isotopes as well as intermediate mass
fragments  with $Z < 9$ have also been measured at incident energy of 35  MeV/A  in 
$\rm ^{70}Zn + ^{70}Zn$ , $\rm ^{64}Zn + ^{64}Zn$, $\rm ^{64}Ni + ^{64}Ni$ reactions and compared to transport calculations. The analysis yielded values denoted by the full squares \cite{isotope}.

The box labelled by FRDM (finite-range droplet model) represents a refinement of the droplet model \cite{mol}, and includes microscopic "shell" effects and the extra binding associated with $N=Z$ nuclei. The FRDM reproduces nuclear binding energies of known nuclei within $0.1\%$, and allows determination of both $S_0=32.5 \pm 0.5$ MeV and $L=70 \pm  15$ MeV.

In Fig.\ref{f:L} the other boxes represent experimental data obtained from measurements
of the neutron skin thickness. In light nuclei with $N \approx Z$ , the neutrons and protons have
similar density distributions. With increasing the neutron number $N$, the radius of the neutron density distribution becomes larger than that of the protons, reflecting the pressure of the symmetry energy. The measurement of the neutron skin thickness is made on the stable nucleus 
$\rm ^{208}Pb$, which has
a closed neutron shell with $N=126$ and a closed proton shell with $Z=82$, hence it is very asymmetric and the neutron skin is very thick. The possibility of measurements of the neutron 
radius in $\rm^{208}Pb$ by the experiment PREX  at Jefferson Laboratory has been widely discussed \cite{horo}.  The experiment should extract the value of the neutron radius in $\rm^{208}Pb$ from parity-violating electron scattering. However, the experimental signature is very small, and the extracted thickness has a
large statistical uncertainty. In the next few years, a second experimental run for PREX could 
reduce this large uncertainty \cite{prex}. 

Recent experimental data obtained by Zenihiro et al.\cite{zeni} on the 
neutron skin thickness of $\rm^{208}Pb$ deduced a value  $\delta R_{np} = \rm 0.211^{+0.054}_{-0.063}~fm$. From the experiments constraints on the symmetry energy were derived, and 
these are plotted in Fig.\ref{f:L} as the short-dashed blue rectangular box labelled 
Pb($\vec p, \vec p$).  

Last, we mention the experimental data on the Pygmy Dipole Resonance (PDR) in very neutron-rich
nuclei such as  $\rm^{68}Ni$  and $\rm^{132}Sn$,  which peaks at excitation energies well below the 
Giant Dipole Resonance (GDR), and exhausts about 5$\%$ of the energy-weighted sum rule \cite{pdr}. In many models it has been found that this percentage is linearly dependent on 
the slope $L$ of the symmetry energy. Carbone et al. \cite{carb} extracted a value of $L=64.8 \pm 15.7$ MeV, and $S_0=32.2 \pm 1.3$ MeV using various models which connect $L$ with the neutron 
skin thickness. Those constraints are shown as a long-dashed rectangle in Fig.\ref{f:L} with the label
PDR.  
 
The predictions of the different EoS are also reported in Fig.\ref{f:L} as full symbols. They are distributed within a large region and they span a wide interval in the values of the parameter $L$. However, the various phenomenological data are at best marginally compatible, and it is difficult to put well definite constraints on the EoS. Tentatively, from these data one can restrict the possible values of the symmetry energy at saturation in a limited interval, approximately $30 < S_0 < 35$ MeV, where all the considered EoS are actually falling. 
 
\begin{figure*}[t]
\includegraphics[scale=0.45,clip]{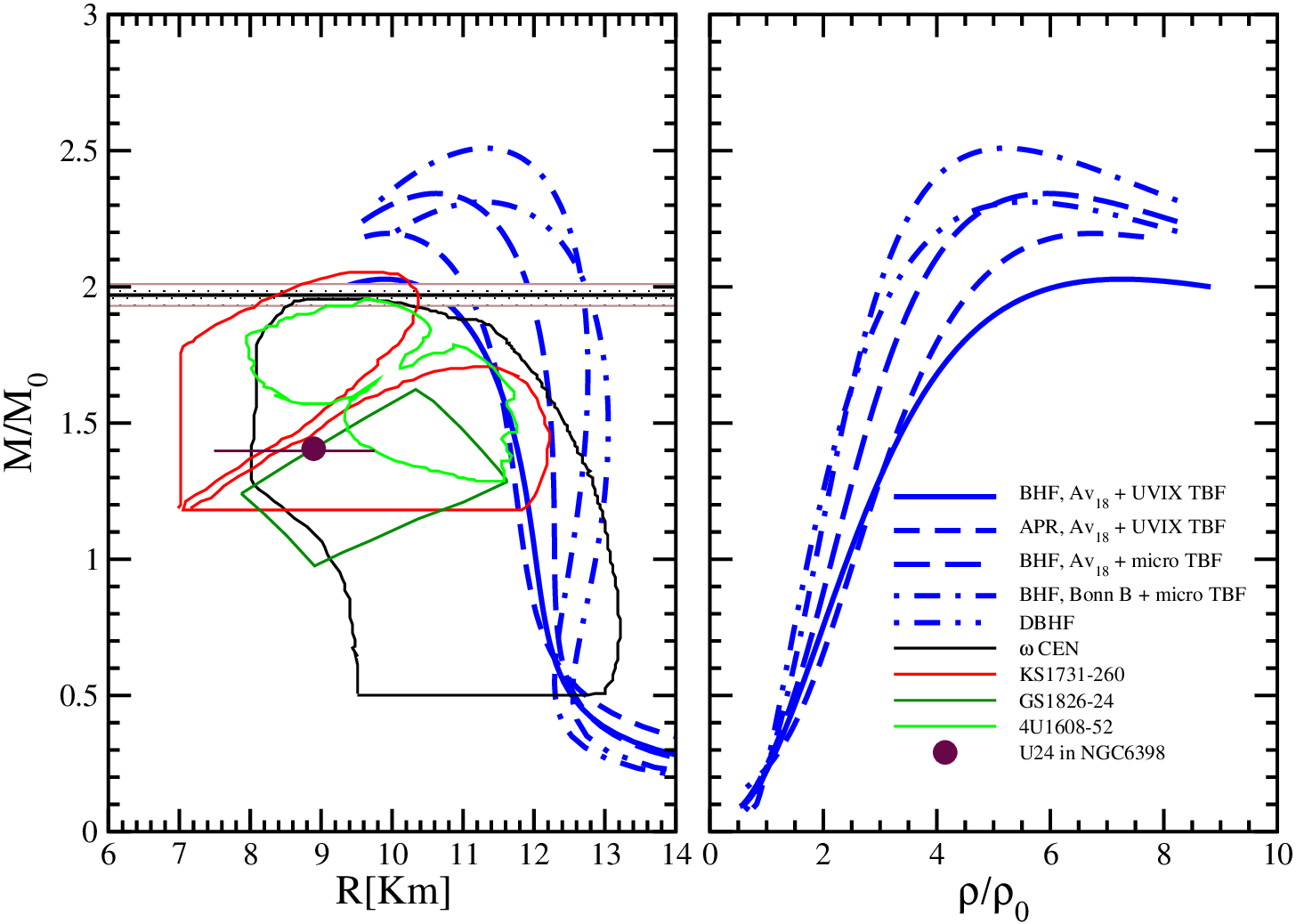}
\caption{(Color online)
The mass-radius (left panel) and the mass-central density (right panel) relations are plotted
for the EoS's discussed. Boxes are boundaries extracted from observations, see ref.\cite{Ozel}.}
\label{f:max}
\end{figure*}
\bigskip

\begin{figure*}[t]
\includegraphics[scale=0.75,clip]{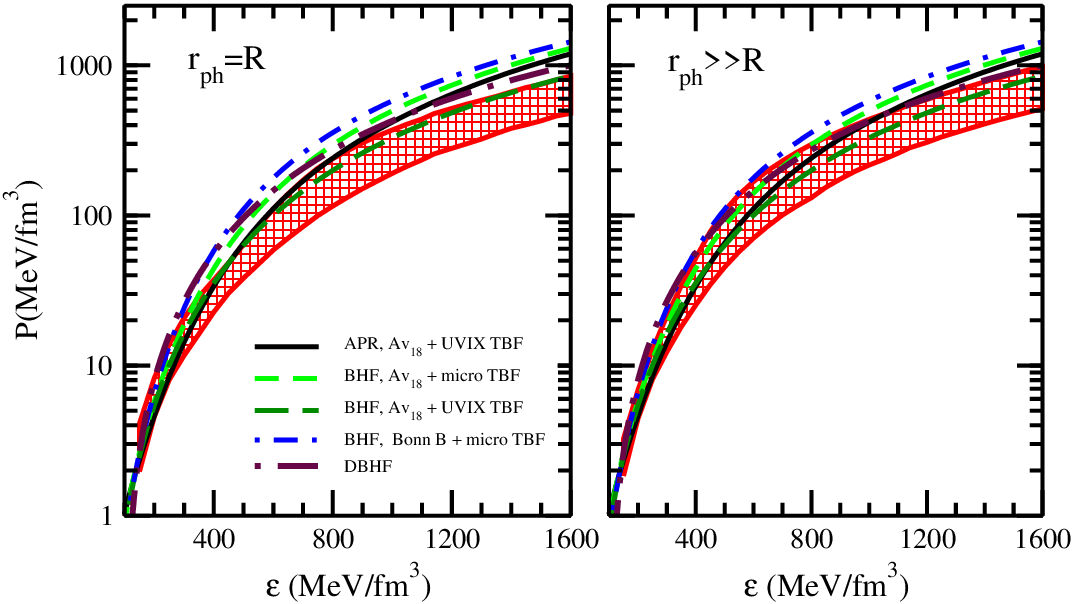}
\caption{(Color online)
Pressure as a function of the mass-energy density in neutron star matter.
The shaded areas  are taken from ref.\cite{baye}. See text for details.}
\label{f:baye}
\end{figure*}

\subsection{Astrophysics \label{astro} }

A neutron star is bound by gravity, and it is kept in hydrostatic equilibrium only by the pressure produced by the compressed nuclear matter. It is then apparent that the nuclear matter EoS is the main medium property that is
relevant in this case, as can be seen in the celebrated Tolman-Oppenheimer-Volkoff \cite{shapiro} equations, valid
for spherically symmetric NS
\begin{eqnarray}
{d P\over d r}&=&- G\, {\epsilon m \over r^2} \left(1 + {P \over \epsilon} \right) \left(1 + {4\pi P r^3\over m } \right) (1 - {2 G m \over r })^{-1} \nonumber \\
{d m\over d r}&=&4\pi r^2 \epsilon 
\label{eq:OV} 
\end{eqnarray}
\noindent where $G$ is the gravitational constant, $P$ the pressure, $\varepsilon$ the energy density, and $r$ the
(relativistic) radius coordinate. To close the equations we need the relation between pressure and density, $P \,
=\, P(\varepsilon)$, i.e. just the EoS. Integrating these equations one gets the mass and radius of the star for each central density. Typical values are
1-2 solar masses ($M_\odot$) and about 10 Km, respectively. This indicates the extremely high density of the object. It turns out that the mass of the NS has a maximum value as a function of radius (or central density), above which the star is unstable against collapse to a black hole. The value of the maximum mass depends on the nuclear EoS, so that the observation of a mass higher than the maximum one allowed by a given EoS simply rules out that EoS. 
The considered microscopic EoS's are compatible with the largest mass observed up to now, that is close to $1.97~\pm 0.04 ~M_{\odot} $  \cite{Demo}. This is clearly shown in Fig.\ref{f:max},
where the mass-radius (left panel) and mass-central density relations (right panel) are plotted
for all the considered EoS's as thick lines. 
It looks unlikely that this value is indeed the largest possible NS mass, and therefore future observational data on NS masses could overcome this limit and strongly constrain the nuclear EoS. 

It would be of course desirable to have some phenomenological data also on the radius of NS. Unfortunately this is quite difficult, but some tentative analysis looks promising \cite{Ozel, guvoz}. 
In Fig.\ref{f:max} a sample of observational data taken from ref.\cite{Ozel} is displayed by closed 
thin lines for different sources, measured in quiescence and from thermonuclear bursts. 
It turns out that the current measurements are consistent with radii in the range 8-12 km and disfavor neutron stars with R $\sim$15 km. Those measurements are consistent with the recent observation of the neutron star in SAX J1748.9-2021, which points to the neutron star radius in the 8-11 km range
\cite{guvoz}.

Additional tentative constraints on the nuclear EoS were obtained in a recent analysis of the data on six NS based on Bayesian statistical framework \cite{baye}. Depending on the hypothesis made on the structure of the NS, the results are slightly different, as shown in Fig.\ref{f:baye}, where 
the quantity $\rm r_{ph}$ is the photosphere radius. In the let panel $r_{ph}$ is comparable to the neutron star radius R, whereas in the right panel a substantial
expansion of the photosphere during an X-ray burst is assumed to occur. The overall allowed region where the EoS's should lie is displayed in Fig.\ref{f:baye} as bounding boxes, where the theoretical EoS's just discussed are also reported as thick lines. Among the different EoS's, only the one calculated with BHF and phenomenological Urbana model appear to be compatible with the extracted observational constraints over the whole
density range. It turns out that other
microscopic EoS do not show the same agreement, in particular the EoS with BHF and microscopic TBF and the DBHF EoS look too repulsive at high density. These boundaries obtained from astrophysical data are complementary to the ones obtained from heavy ion reactions, and illustrated in the previous subsection. In fact, in heavy ion
collisions the tested matter is essentially symmetric, while in NS the matter is highly asymmetric. Considered together, the two types of constraints probe the density dependence of the symmetry energy.

\begin{figure}[t]
\includegraphics[scale=0.35,clip]{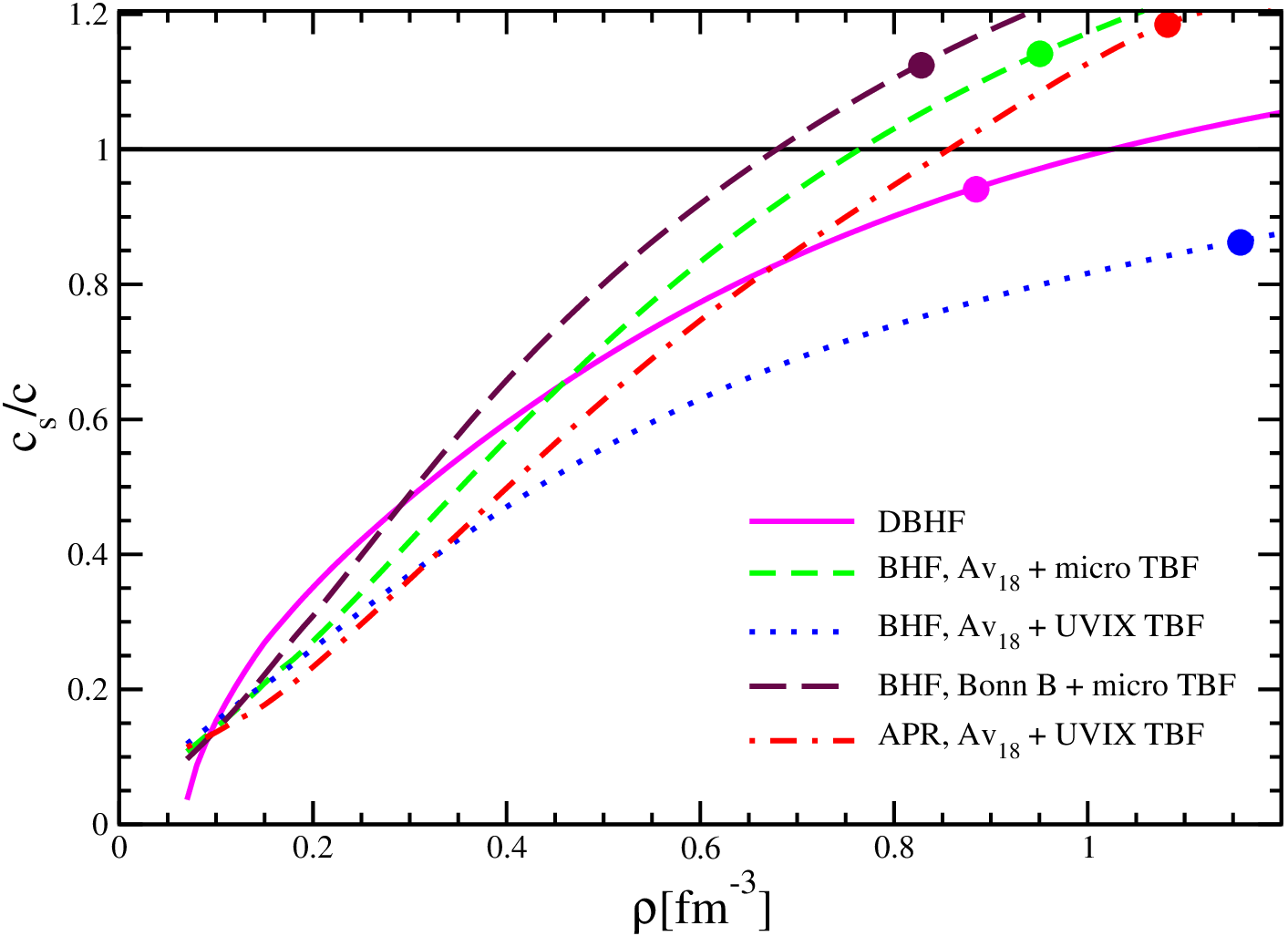}
\caption{(Color online)
The speed of sound is plotted as function of the nucleon density for the EoS's discussed
in the text. The dots mark the central density for the maximum mass of a neutron star.}
\label{f:cs}
\end{figure}

\begin{figure}[t]
\includegraphics[scale=0.35,clip]{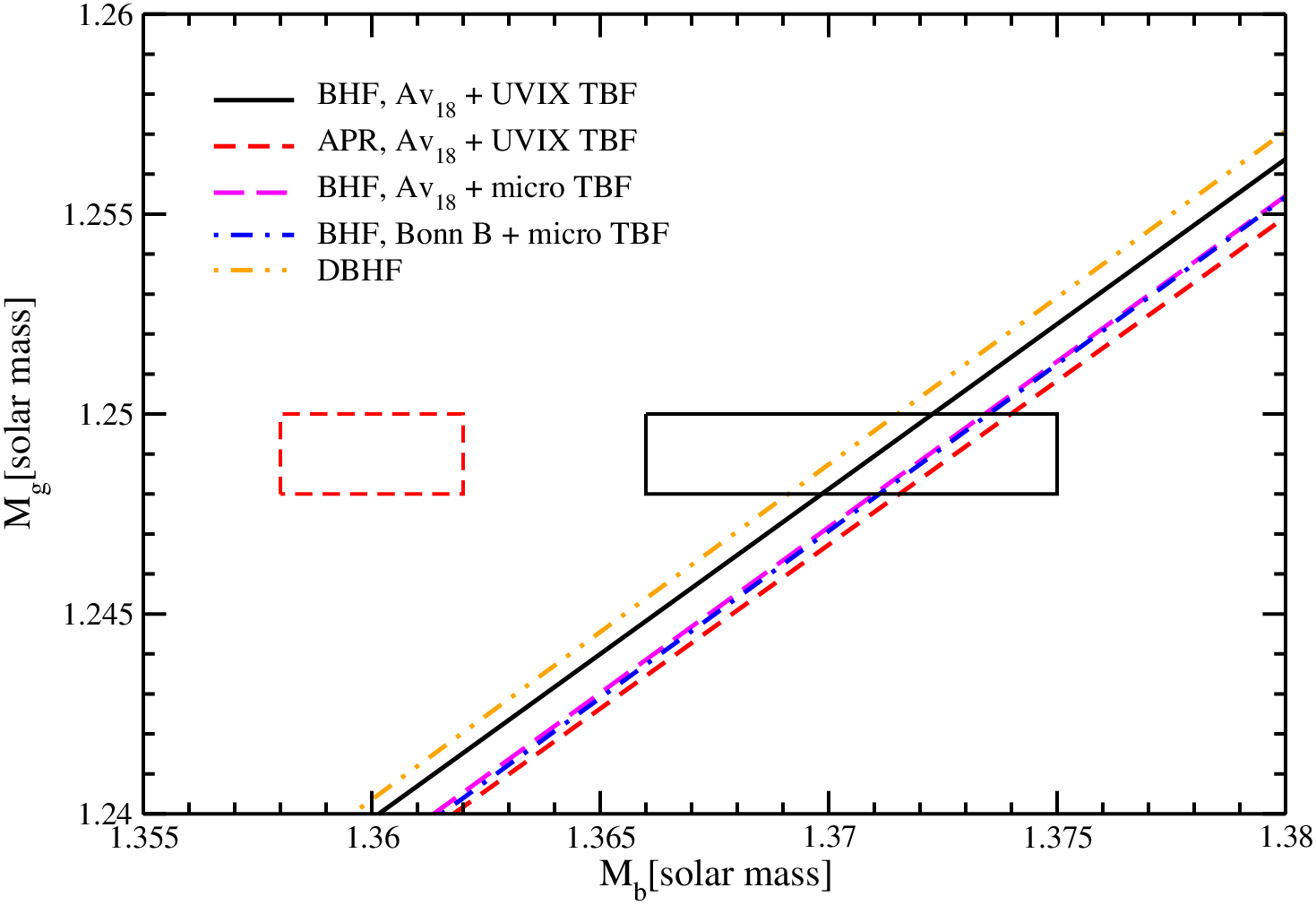}
\caption{(Color online)
The gravitational mass is plotted as a function of the baryon mass
for the EoS's discussed in the text. The boxes indicate the boundaries 
coming from the simulations.}
\label{f:mgmb}
\end{figure}

\par
In relation to the high density region of the nuclear EoS, an additional test is on the speed of sound $c_S$, that is required to be smaller than the speed of light $c$ (causality condition). The speed of sound is directly connected with the incompressibility and the energy density, according to the relativistic expression 
\begin{equation}
\frac{c_S}{c} \, =\, \sqrt{\frac{dP}{d\epsilon}}
\end{equation}
As such, it depends on the matter energy density and asymmetry.
In Fig.\ref{f:cs} we plot the speed of sound in units of $c$  as a function of the density of the
NS matter, according to each EoS. In the figure the full points on each curve indicate the central density of the calculated NS maximum mass for the given EoS. At large enough density most of the EoS's show a superluminal speed of sound. One finds that also the DBHF EoS shows a superluminal behavior. This is not completely surprising, since the DBHF approach is actually based on the three-dimensional reduction of the original four-dimensional Bethe-Salpeter equation, and therefore it is not fully relativistic.

A further additional constraint on the neutron star EoS is provided by the observation of the
double pulsar J0737-3039, and the interpretation given by Podsiadlowski \cite{Pod}.
In fact, the gravitational mass of Pulsar B is very precisely known $\rm M_G = 1.249 \pm 0.001 \, M_{\odot}$, whereas  estimates of the baryonic mass depend upon its detailed mode of formation. 
As modelled by Podsiadlowski et al., if  the pulsar B was formed from a white dwarf with an O-Ne-Mg core in an electron-capture supernova, assuming no or negligible loss of baryonic mass during the collapse, the newly born neutron star will have the same baryonic mass as the precollapse core of the progenitor star, i.e. $\rm M_B \simeq 1.366-1.375 \, M_\odot$. This result is displayed in Fig.\ref{f:mgmb}
as a black box.  Though, taking into account the uncertainty in the EoS and the small mass loss during the collapse,  Kitaura et al. \cite{kita} made another simulation which gave  
$\rm M_B = 1.360 \pm 0.002 \, M_\odot$, which is shown in Fig.\ref{f:mgmb} by the dashed red box.
 We have calculated for each neutron star matter EoS the relation between the gravitational
 and baryonic mass, and these are displayed in Fig.\ref{f:mgmb} by the straight 
 curves. We notice that the results of all microscopic EoS's agree very well
 with the result of Podsiadlowski, at variance with the calculations based on the phenomenological  Skyrme forces discussed in ref.\cite{dutra}, where agreement was found with the result of 
 Kitaura et al. \cite{kita}, which assumed small mass loss during the collapse.

\section{Conclusions}
\label{s:end}
We have presented a systematic confrontation of the nuclear Equation of
State, obtained  within different microscopic many-body methods, with the
available constraints coming from phenomenology.
The latter are extracted from laboratory experiments as well as from
astrophysical observations. Both nuclear structure and heavy ion
collisions data were considered, along the same lines of the analysis on
the Skyrme forces reported in ref. \cite{dutra}. Astrophysical observational
data included the measures of NS masses, some hints on the radius-mass
relation from ref. \cite{Ozel,guvoz} and the constraints on the EoS presented in
ref. \cite{baye}, obtained from the analysis of transient phenomena in six
NS's. Some theoretical constraints, as the requirement of a sub-luminal
speed of sound, were also considered.  If one takes literally all
the constraints, among the considered microscopic EoS only one passes all
the tests. However, these phenomenological constraints are affected by
uncertainties, which are difficult to estimate quantitatively on a firm
basis. The conclusion that one can draw from this analysis is twofold.
Firstly, despite the differences between the considered microscopic EoS,
their overall predictions do not show major discrepancies with one another
as well as with the phenomenological constraints. In other words, the
calculated microscopic nuclear EoS, based on different many-body methods,
are not in contradiction with the phenomenological constraints. This is
not at all an obvious result. Secondly, from the analysis it appears that
one can explain reasonably well all the data with a microscopic EoS that
includes only nucleonic degrees of freedom, in particular no exotic
components in NS are needed. On one hand, it is likely that this
conclusion will be disproved in the future with the expected new data, in
particular on the NS maximum mass. Indeed, if the evidence for a NS mass
of 2.5 solar mass \cite{romani} will be confirmed, this will rule out most of the considered microscopic
EoS and will introduce a serious and fundamental issue in the physics of NS.
On the other hand, it has been shown
that exotic matter like hyperons or quarks should appear in NS \cite{sch98}, which will
strongly affect the EoS and challenge the theory of high density nuclear
matter. In the future one can expect that the interplay between theory and
observations will continue to play a major role
in the worldwide effort of determining the nuclear EoS.
\bigskip

\section*{Acknowledgments}

We warmly thank P. Danielewicz and J. Lee for providing us the data on the symmetry energy 
shown in Fig.\ref{f:esym}. Moreover we acknowledge H.-J. Schulze for the data concerning the EoS calculated in the BHF approach with Bonn B NN potential and microscopic TBF's,
and C. Fuchs for the DBHF data set.


\end{document}